\documentclass[preprint,showpacs,preprintnumbers,amsmath,amssymb]{revtex4}

\newcommand{\RR}{\mathrm{R}}
\newcommand{\TT}{\mathrm{T}}

\newcommand{\HH}{\mathrm{H}}

\newcommand{\Id}{\mathrm{I_d}}
\usepackage{color}
\definecolor{black}{gray}{0.65}
\usepackage{graphicx}	
\usepackage{dcolumn}	
\usepackage{bm}		
\usepackage{psfrag} 
\begin{document}

\preprint{APS/123-QED}

\title{Multiple Scattering and Visco-Thermal Effects on 2D Phononic Crystal\\} 
\author{A.~Duclos, D.~Lafarge, and V.~Pagneux}
\address{Laboratoire d'Acoustique de l'Universit{\'e} du Maine, UMR-CNRS 6613, Universit{\'e} du Maine, Av. Olivier Messiaen 72085 Le Mans cedex 9, France}

\date{\today}
\begin{abstract}
In this paper, we are interested in the transition between regimes where either visco-thermal or multiple scattering effects dominate for the propagation of acoustic waves through a 2D regular square array of rigid cylinders embedded in air. An extension of the numerical method using Schl{\"o}milch series is performed in order to account for visco-thermal losses. Comparison with experimental data and results from classical homogenization theory allows to study the transition between a low frequency limit (where viscous and thermal effects dominate) and a high frequency regime (where multiple scattering effects become predominant). For this particular geometry, a large frequency domain where visco-thermal and multiple scattering effects coexist is found.
%
\end{abstract}
\pacs{Valid PACS appear here}
\maketitle


\section{Introduction}

{A precise description of sound propagation and absorption in phononic crystal for a large frequency domain is of great interest in order to study transition between low and high frequency}. At low frequencies, i.e. when the viscous and thermal skin depths are comparable to the {crystal characteristic dimension}, important dissipation effects occur. These effects are well described by the long wavelength homogenization theories~\cite{John87,Burr85,Sheng88,Alla93,Lafa97,Smeul92}. Simple fluid dynamics problems are to be solved: a Stokes unsteady fluid flow problem for the velocity, and a Fourier unsteady heat conduction problem for the temperature. Studies describing the propagation by the homogenization theory can be found in literature~\cite{Cort02,Cort03}.
When the frequency increases enough, such separation no longer is possible and the problem transforms into a full wave propagation problem with multiple scattering effects becoming predominent. These effects may lead to band gaps (frequency regions where the propagation is forbidden). Different existing numerical methods allowing to determine these band gaps can be found in the literature~\cite{Pere93,Vass01,Sanch98,Kush94,Kush97,McPh05}. For the purpose of studying the transition between low and high frequency regime we first extend the numerical method using Schl{\"o}milch series~\cite{Twer61,Twer63,Lint06} to include visco-thermal effects. As a result, a numerical method is obtained which takes into account viscous, thermal and multiple scattering effects. Then, a comparison with experimental data (high frequencies) on one hand and with homogenization theory results (low frequencies) on the other hand is performed. Finally the transition between low and high frequencies is discussed.


\section{Theory}
\begin{figure}
\includegraphics [width=8.5cm] {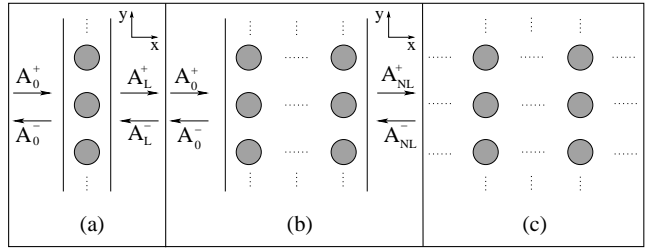}
\caption{\label{fig_problem} Geometries under consideration. (a) one row containing an infinite number of cylinders, (b) n-rows containing an infinite number of cylinders, and (c) infinite phononic crystal.}
\end{figure}

We consider sound propagation and absorption through a 2D periodic square arrangement of rigid cylinders embedded in air for a large frequency domain. This domain includes important multiple scattering as well as visco-thermal effects. The three geometries represented on figure~\ref{fig_problem} are considered. The first problem is a row containing an infinite number of cylinders periodically spaced according to $y$ axis and insonified by (right and left) ingoing plane waves. The second one is an arrangement of $N$ rows of cylinders with the same periodicity according to $x$ and $y$ axis and subject to the same ingoing plane waves. In a last problem, we consider the case of an infinite periodic medium called ``phononic crystal''.
For the cases (a) and (b), we are interested in the reflection and transmission properties of the samples, and for the infinite phononic crystal (c), we study the dispersion relation of the Floquet-Boch modes. The solution of the problems (b) and (c) is based on extensions of the solution of the problem (a). In order to account for visco-thermal and multiple scattering effects, two descriptions are introduced. Firstly, we consider an equivalent surface admittance surrounding each cylinder (method A). Using this concept, previously described by Tournat et al.~\cite{Tournat04}, weak dissipation localized in the cylinder boundary layers can be considered. Secondly, a complete description allowing for losses (localized or not) in all the fluid is presented (method B). Analytically, the two problems are formulated in terms of the displacement $\vec{u}$ and the (excess) temperature $\tau$ fields. In the first case (method A), these fields derive from an acoustic scalar potential $\phi_a$,
\begin{eqnarray}
\left\{\begin{array}{ll}
\vec{u}=\overrightarrow{\nabla} \displaystyle \ \phi_a \\
\tau= \displaystyle \frac{\omega^2}{C_p} \phi_a
\label{utau_ad}
\end{array} \quad , \right.
\end{eqnarray}
which satisfies the 2D Helmholtz equation~\cite{Mors68} (convention $e^{-i\omega t}$),
\begin{eqnarray}
(\Delta+k_a^2) \phi_a&=&0\quad,\quad k_a^2=\left( \displaystyle {\omega}/{c_0} \right)^2 \,
\end{eqnarray}
where $\omega$ is the angular frequency and $c_0$ is the adiabatic sound speed. The visco-thermal effects occuring in the boundary layers are considered in the following boundary condition set at cylinder surfaces,
\begin{eqnarray}
\left(\partial_n +i k_a {\beta}\right)\phi_a=0\ ,\label{boundary_ad}
\end{eqnarray}
where $\partial_n$ is the normal derivative and $\beta$ is the equivalent surface admittance operator which depends notably on fluid properties (kinematic viscosity and thermal diffusivity). Expression of this operator (in the limit of small boundary layers) can be found in~\cite{Tournat04} for the 2D cylindrical case.

In the second case (method B), the displacement and temperature fields can be related~(\cite{Pier81}) to three potentials : acoustic $\phi_a$, entropic $\phi_h$ and vorticity $\vec{\psi}$,
\begin{eqnarray}
\left\{\begin{array}{ll}
\vec{u}=\overrightarrow{\nabla} \displaystyle \left(\phi_a+\phi_h\right)+ \overrightarrow{\nabla}\wedge \vec{\psi} \\
\tau =  \displaystyle \frac{\omega^2}{C_p} \phi_a - \frac{i \omega}{\beta_0 \nu_h} \phi_h \label{3-pot}
\end{array} \quad , \right.
\end{eqnarray}
where $\beta_0$ is the coefficient of thermal expansion, $T_0$ the temperature, $C_p$ the specific heat at constant pressure and $\nu_h$ the thermal diffusivity. All these parameters depend on the fluid nature. The three potentials (we set $\vec{\psi}=\phi_v\vec{e_z}$) obey to the 2D Helmholtz  equations~(\cite{Pier81}),
\begin{eqnarray}\left\{ \begin{array}{lll}
(\Delta+k_a^2)\phi_a&=&0\quad,\quad k_a^2= \left( \displaystyle {\omega}/{c_0} \right)^2\\
(\Delta+k_h^2)\phi_h&=&0\quad,\quad k_h^2=\displaystyle {i \omega}/{\nu_h} \\ 
(\Delta+k_v^2)\phi_v&=&0\quad,\quad k_v^2=\displaystyle {i \omega}/{\nu_v}
\end{array}\right.\quad ,\label{k_ahv}
\end{eqnarray}
where $\nu_v$ is the kinematic viscosity. The real wavenumber $k_a$ is related to acoustic propagation while the complex wavenumbers $k_v$ and $k_h$ are related to viscous and thermal waves respectively. The three potentials are coupled at the cylinders boundaries by the following conditions,
\begin{eqnarray}
\left\{\begin{array}{ll} \vec{u}=\vec{0} \\
\tau=0
\end{array} \right.\ ,\label{C-L}
\end{eqnarray}
cancelling displacement and temperature fields. 

Having in view the determination of the Floquet-Bloch modes (in the $x$ direction of the phononic cristal), we will in methods A and B expand the potentials in terms of right and left going plane waves,  
\begin{eqnarray}
\phi_{pot}(\vec{r})=\displaystyle \sum_{n=-\infty}^{\infty}  \left( A_{pot,n}^{+}e^{i\vec{k}_{pot,n}.\vec{r}}+A_{pot,n}^{-}e^{-i\vec{k}_{pot,n}.\vec{r}}\right) , \label{pot_cart_e}
\end{eqnarray}

where the ``${pot}$'' index is either ${a}$ (for the acoustic potential), ${h}$ (for the entropic potential) or ${v}$ (for the vorticity potential), and the index $n$ determines the possible wavevectors through ($\vec{k}_{pot,n}^2=k_{pot}^2$) and the condition $(\vec{k}_{pot,n})_y L=2\pi n$ which must be satisfied in order to respect the $y$ periodicity. The amplitudes of all potentials are merged into coefficients $A$ in figure~(\ref{fig_problem}). To obtain the reflection and transmission properties of one row (case (a)), the scattering matrix, which relates the outgoing waves to the ingoing ones, is introduced,
\begin{eqnarray}
\left( \begin{array}{c}
\vec{A}_0^{~-} \\
\vec{A}_L^{~+}
\end{array} \right)
=\left[ \begin{array}{cc}
{[\TT]} & {[\RR]}\\
{[\RR]} & {[\TT]}
\end{array} \right]
\left( \begin{array}{cc}
\vec{A}_L^{~-} \\
\vec{A}_0^{~+}
\end{array} \right)	\ ,\label{mat_diff2}
\end{eqnarray}
where $[\RR]$ and $[\TT]$ are the reflection and transmission matrices. To determine theirs elements $r_{nn'}^{pot\ pot'}$ and $t_{nn'}^{pot\ pot'}$, the outgoing potentials are expressed in two different forms, on each side of the row for a given ingoing potential $\phi_{ext}(\vec{r},n',pot')=e^{i\vec{k}_{pot',n'}.\vec{r}}$.  The outgoing potentials can be interpreted as the reflected and transmitted fields by the row,
\begin{eqnarray}
\left\{\begin{array}{lll}
\phi_{pot,0}^{R}(\vec{r})&=&\displaystyle \sum_{n=-\infty}^{\infty}r_{nn'}^{pot\ pot'}e^{-i\vec{k}_{pot,n}.\vec{r}} \\
\phi_{pot,L}^{T}(\vec{r})&=&\displaystyle \sum_{n=-\infty}^{\infty}t_{nn'}^{pot\ pot'}e^{i\vec{k}_{pot,n}.\vec{r}}
\end{array}\right. \ . \label{pot_RT}
\end{eqnarray}
These can also be interpreted in term of the potential fields scattered by all cylinders belonging to the row,
\begin{widetext}
\begin{eqnarray}
\left\{\begin{array}{lll}
\phi_{pot,0}^{R}(\vec{r})&=&\displaystyle \sum_{n=-\infty}^{\infty}\sum_{j=-\infty}^{\infty}
C_{nn'}^{pot\ pot'}i^m\HH_m(k_{pot}|\vec{r}-\vec{r_j}|)e^{im\Phi_{\vec{r}-\vec{r_j}}}\\
\phi_{pot,0}^{T}(\vec{r})&=&\displaystyle \sum_{n=-\infty}^{\infty}\sum_{j=-\infty}^{\infty} D_{nn'}^{pot\ pot'}i^m\HH_m(k_{pot}|\vec{r}-\vec{r_j}|)e^{im\Phi_{\vec{r}-\vec{r_j}}} + \phi_{ext}(\vec{r},n',pot')
\end{array}\right. \ , \label{pot_scatt}
\end{eqnarray}
\end{widetext}
where the coefficients $C_{nn'}^{pot\ pot'}$ and $D_{nn'}^{pot\ pot'}$ are independent of the scatterer (their determination is performed numerically according to~\cite{Chen01-1} for instance). At this point, we must estimate Schl{\"o}milch series of the form $\sum_{j=-\infty}^{\infty}i^m\HH_m(k_{pot}|\vec{r}-\vec{r_j}|)e^{im\Phi_{\vec{r}-\vec{r_j}}}$. For the usual case $k_{pot}=k_a$ defined in relation~(\ref{k_ahv}) with $k_a\in \mathbb{R}$, and due to the slow convergence of Schl{\"o}milch series, these sums are developed in the form~\cite{Twer61,Twer63,Lint06} which present a better convergence. For the more original cases $k_{pot}=k_h$ and $k_{pot}=k_v$ also defined in relation~(\ref{k_ahv}) and having non zero $Im(k_h)$ and $Im(k_v)$, the Schl{\"o}milch series are naturally convergent. Then an identification between the two couples of equations~(\ref{pot_RT}) and (\ref{pot_scatt}) lead to the required reflection and transmission coefficients. Consequently the reflection and transmission properties of one row (case (a)) are entirely determined.

Now we consider the case (b) of $N$ rows (a) separated by the distance $L$. The outgoing and incoming wave amplitudes are related by a new scattering matrix,
\begin{eqnarray}
\left( \begin{array}{c}
\vec{A}_0^{~-} \\
\vec{A}_{NL}^{~+}
\end{array} \right)
=\left[ \begin{array}{cc}
{[\TT_N]} & {[\RR_N]}\\
{[\RR_N]} & {[\TT_N]}
\end{array} \right]
\left( \begin{array}{cc}
\vec{A}_{NL}^{~-} \\
\vec{A}_0^{~+}
\end{array} \right)	\ , \label{mat_diff3}
\end{eqnarray}
where $[\RR_N]$ and $[\TT_N]$ are the reflection and transmission matrices for $N$ rows. These two matrices can be related to $[\RR]$ and $[\TT]$ using the following recurrence relations,
\begin{eqnarray}
\left \{ \begin{array}{lllll}
[\TT_{N}]=[\TT]\Big([\Id]-[\RR_{N-1}][\RR]\Big)^{-1}[\TT_{N-1}] \\
{}[\RR_{N}]=[\RR]+[\TT]\Big([\Id]-[\RR_{N-1}][\RR]\Big)^{-1}[\RR_{N-1}][\TT] 
\end{array} \right. ,\label{compo}
\end{eqnarray}
where $[\Id]$ is the identity matrix. Using these relations, the reflection and transmission properties of $N$ rows are simply deduced from the two known matrices $[\RR]$ and $[\TT]$.

The phononic crystal (case (c)) is interpreted as a superposition of an infinite number of rows. Taking an arbitrary row, the scattering matrix concept (relation~(\ref{mat_diff2})) remains valid. Moreover,  the potentials obey the Bloch condition~\cite{Bril46},
\begin{eqnarray}
\left( \begin{array}{ccc}
\vec{A}_L^{~+} \\
\vec{A}_L^{~-}
\end{array} \right)
=e^{ik_BL}
\left( \begin{array}{ccc}
\vec{A}_0^{~+} \\
\vec{A}_0^{~-}
\end{array} \right), \label{mat_bloch}
\end{eqnarray}
where the $k_B$ are the Bloch wavenumbers to determine. The simultaneous use of the two relations~(\ref{mat_diff2}) and (\ref{mat_bloch}) leads to the following eigenvalue problem,
\begin{eqnarray}
\left[ \begin{array}{ccc}
{[\TT]} & {[\RR]} \\
{[0]} & {[\Id]}
\end{array} \right]
\left( \begin{array}{ccc}
\vec{A}_0^{~+} \\
\vec{A}_L^{~-}
\end{array} \right)
=e^{ik_BL}
\left[ \begin{array}{ccc}
{[\Id]} &  {[0]}\\
{[\RR]} & {[\TT]}
\end{array} \right]
\left( \begin{array}{ccc}
\vec{A}_0^{~+} \\
\vec{A}_L^{~-}
\end{array} \right) ,\label{propre}
\end{eqnarray}
where $[0]$ is the zero matrix. As the reflection and transmission matrices $[\RR]$ and $[\TT]$ are already known, the dispersion relation giving the solution for the Bloch wavenumbers $k_B$ can be solved numerically.


\section{Experimental and theoretical results}
\begin{figure}[!]
\psfrag{kL/pi}{\small $k_aL/\pi$}
\includegraphics[width=8.5cm]{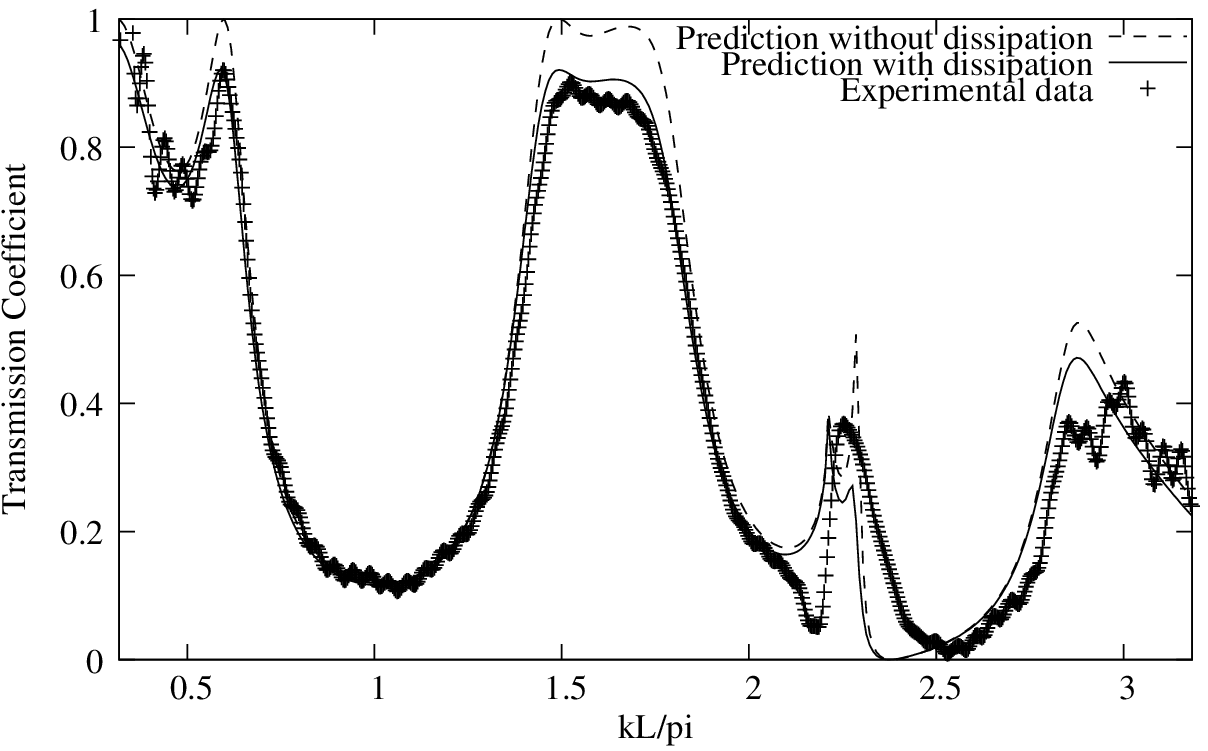}\\
\includegraphics[width=8.5cm]{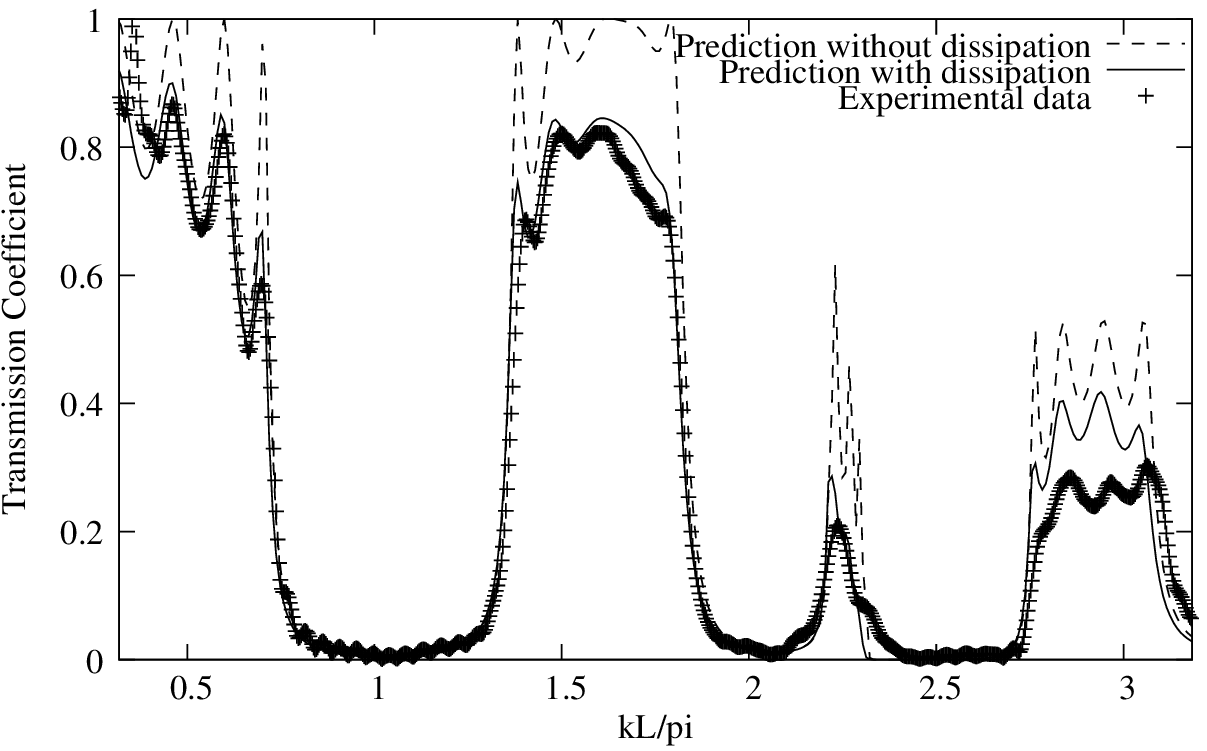}
\caption{\label{fig_rangees} Transmission coefficients of two samples containing 3 rows (upper) and 6 rows (lower) versus reduce frequency $k_aL/\pi$ (corresponding to a frequency range going from 20~kHz to 200~kHz). Cylinder radius is R=1~mm and lattice periodicity is L=3~mm (thus porosity is $\Phi=0.65$). We have reported the theoretical predictions with ($-$) and without ($--$) dissipation effects and the experimental data ($+$).}
\end{figure}

A first comparison is performed at high frequencies (acoustic wavelength comparable or less than the lattice parameter $L$, $k_aL/\pi \geq 1$) corresponding to a regime governed by important multiple scattering effects. The two models described above allow to compute the scattering matrix~(\ref{mat_diff2}) of one row and then, using equation~(\ref{compo}), to predict the transmission coefficient of a sample containing a finite number of rows at normal incidence. This transmission coefficient can be experimentally obtained. Here, two samples of aluminium rods embedded in air in square configuration (as presented in figure~\ref{fig_problem} (b)) are used. The radius of the cylinders is fixed at R=1~mm. The lattice parameter of the sample is $L$=3~mm. The two samples contain respectively  3 and 6 rows. Each of them is 100 cylinders wide. The experiment is carried out at ultrasonic frequencies using a couple of broadband transducers (centered on 100~kHz) with radius 20~mm. The experimental results are reported on figure~\ref{fig_rangees}. When the number of rows increases, the transmission coefficient tends towards zero inside frequency domains corresponding to band gaps defined for infinite phononic crystals. We have also reported the predictions of the two models A and B, which are undistinguishable, and the prediction for the ideal fluid (i.e. $\beta=0$ in~ (\ref{boundary_ad})). The band gap locations are well described in all cases because they are essentially due to multiple scattering effects only. However, outside these bands, an accurate prediction of the transmission coefficients needs taking into account the visco-thermal losses. The two visco-thermal descriptions lead to the same predictions because the dissipation effects occuring during the propagation are entirely localized in the small boundary layers. Discrepancies between the ideal and visco-thermal fluid models are solely due to these weak losses.

\begin{figure}[!]
\psfrag{Re [V]}{\small {Re $[V]$ $(m.s^{-1})$}}
\psfrag{k_aL/pi}{\small $k_aL/\pi$}
\includegraphics[width=8.5cm]{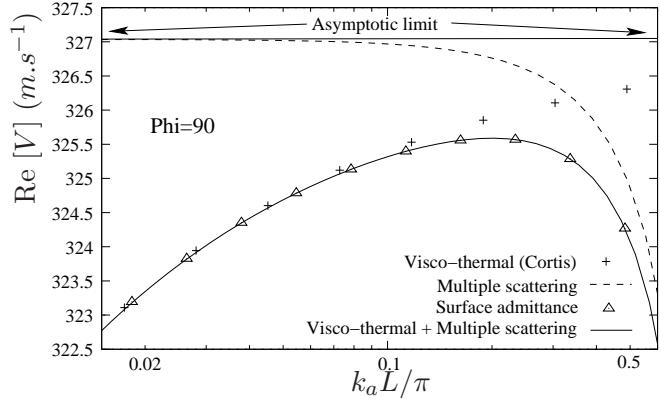}
\caption{\label{fig_V} Real part of phase velocity versus reduce frequency $k_aL/\pi$. Phononic crystal made up of 1~mm radius cylinders (porosity is $\Phi=0.90$). We have reported the theoretical predictions considering visco-thermal effects ($+$) obtained by Cortis, multiple scattering ($--$), equivalent surface admittance concept  ($\Delta$) and visco-thermal as well as multiple scattering effects ($-$). The two visco-thermal models considering multiple scattering effects can describe the transition between low and high frequency regions. Indeed, in this present case, weak dissipation effects are localized at cylinder boundary layers.}
\end{figure}

A comparison between our computations and the low frequency (i.e. long wavelength) model is now performed when the dissipation effects become more important. Previous numerical studies based on homogenization theory (limited to leading order terms) and using finite element methods to solve the problem of long wavelength sound absorption in 2D phononic crystal (figure~\ref{fig_problem} case (c)) have been performed by Cortis~\cite{Cort02}. The principle of these long wavelength simulations is to solve the unsteady Stokes equation for velocity and the unsteady Fourier equation for temperature. It exploits in consistent manner the fact that (to leading order) the velocity field becomes divergence-free at the pore scale (for the purpose of evaluating viscous effects) and the pressure field becomes uniform (for the purpose of evaluating thermal effects). In order to compare these existing homogenization results to those given by the two visco-thermal models including multiple scattering, we compare the real part of our complex velocity $V$ associated to the phononic crystal with that given by Cortis results. This quantity can be expressed from the real part of the Bloch wavenumber $k_B$, solution of the eigenvalue problem~(\ref{propre}), by the relation~:~$V={\omega}/{Re[k_B]}$. The evolution of velocity versus reduced frequency ($k_aL/\pi$) is presented on figure~\ref{fig_V} for a first sample of 0.90 porosity containing 1~mm cylinders embedded in air. Once again, we have reported the results for the ideal fluid (considering the multiple scattering effects only). At the low frequency limit, the two visco-thermal models tends towards Cortis results. Consequently and for the geometrical dimension considered here, our two models successfully fulfill the two asymptotic limits. This is because in the low frequency regime considered here, the dissipation effects are still localized in small boundary layers. When the frequency increases, multiple scattering effects emerge and, as expected, the (leading order terms) classical homogenization results are no more able to predict the correct velocity. Here, the real part of the velocity decreases and tends towards zero quickly due to the first band gap where the propagation is forbidden. At this point, we can precisely study the transition between a long wavelentgh visco-thermal description of sound propagation and the emergence of multiple scattering effects. The limit of the classical homogenization theory for a simple 2D phononic crystal of 0.90 porosity case appears near $k_aL/\pi=0.2$. The same limit was found for other porosities decreasing from 0.90 to 0.50. Description of sound absorption using classical homogenization theory is not accurate beyond this frequency limit. In the presented case, the dissipation effects are very weak ($1\%$ modification on phase velocity) and localized in very small boundary layers. Then, the transition description can be performed by the equivalent surface admittance concept (method A) as well as the complete description (method B).

\begin{figure}[!]
\psfrag{Re [V]}{\small {Re $[V]$ $(m.s^{-1})$}}
\psfrag{k_aL/pi}{\small $k_aL/\pi$}
\includegraphics[width=8.5cm]{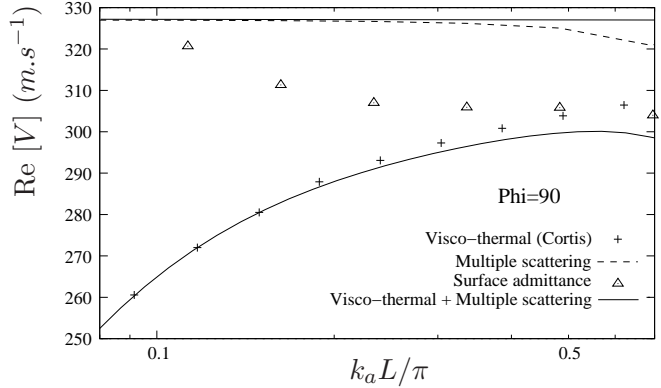}
\caption{\label{fig_Vµ} Real part of phase velocity versus reduce frequency $k_aL/\pi$. Phononic crystal made up of $1~\mu m$ radius cylinders. In this second geometry, the transition can only be described considering important visco-thermal effects not localized at cylinder boundary layers.}
\end{figure}

{Now, we translate the frequency values and we consider a 1~$\mu$m radius cylinders configuration. This scale law allows us to consider the same multiple scattering effects but more important dissipation. Associated results are reported} on figure~\ref{fig_Vµ} in the transition domain where both visco-thermal and multiple scattering effects are significant. {As expected,} the observed limit for classical homogenization theory, due to multiple scattering emergence, remains unchanged. In this second case, our equivalent surface admittance concept cannot describe the transition due to important visco-thermal effects (about $20\%$ modification on phase velocity). For this geometry, the transition shall be described by considering the complete multiple scattering calculations.

Besides, we know that in the high frequency limit, the velocity $V(\omega)$ given by the homogenization theory reaches an asymptotic value which is related to the tortuosity $\alpha_\infty$~\cite{John87} (geometrical characteristic of porous media) with the following relation,
\begin{eqnarray}
\lim_{\omega \to \infty} V(\omega) = \frac{c_0}{\sqrt {\alpha_\infty} } \ .
\end{eqnarray}
{This high frequency limit in view of long wavelength theory is just that matched by the ideal fluid results, in view of multiple scattering theory, in the low frequency limit (absence of multiple scattering effects)}. Corresponding values are indicated on figures~\ref{fig_V} and~\ref{fig_Vµ} by the top horizontal line. The tortuosity values obtained with the ideal fluid model (at the low frequency limit) are in very good agreement with those from literature~\cite{Perr79,Cort02}. However, these values are never reached when considering both visco-thermal and multiple scattering effects as illustrated on figures~\ref{fig_V} and \ref{fig_Vµ}. This observation confirms the existence of a large frequency domain (from $k_aL/\pi=0.2$ for this example) where viscous, thermal and multiple scattering effects coexist. In this region, sound propagation and absorption can only be described by taking into account all these effects together.

\section{Conclusion}
In summary, we have extended an existing model describing the multiple scattering effects on 2D phononic crystal (using Schl{\"o}milch series) by including viscous and thermal effects. The main results consists in the existence of a large frequency domain where both multiple scattering and visco-thermal effects shall be taken into account. For an usual millimetric geometry surrounded by air, dissipation effects are very weak in the transition domain. Thus the description can be performed using the simple equivalent surface admittance concept considered in~\cite{Tournat04} (including the visco-thermal effects in the small boundary layers near the cylinders). However, for a micrometric geometry, this simple description becomes inaccurate due to the fact that the boundary layers are not small. In this paper, this case was successfully described by a general complete multiple scattering calculation. However, the studied transition will also be reproduced by an extended addmittance approach (removing the assumption of small boundary layers) as long as the viscous and thermal boundary layers are not overlapping between the different cylinders. Furthermore, we have shown that we can precisely recover the high frequency limit of the classical homogenization results with our complete multiple scattering calculation in the given 2D phononic crystal. 


\end{document}